\documentclass{scrartcl}

\usepackage{hyperref}
\usepackage{amsfonts}
\usepackage{amsmath}
\usepackage{amssymb}
\usepackage{latexsym}
\usepackage{graphicx}
 
\usepackage{yhmath}
\usepackage{amsmath}
\usepackage{empheq}

\newcommand\no{\nonumber\\{}}

\newcommand\eqnb{\begin{eqnarray}}
\newcommand\eqne{\end{eqnarray}}

\newcommand{\mf}{\mathfrak}

\newcommand{\C}{\ensuremath{\mathbb{C}}}
\newcommand{\R}{\ensuremath{\mathbb{R}}}
\newcommand{\Z}{\ensuremath{\mathbb{Z}}}

\newcommand\void[1]{}   
\newcommand{\Tr}{{\mathrm{Tr}}}

\usepackage{latexsym}

\numberwithin{equation}{section}

\def\cF{\mathcal{F}}

\def\cT{\mathcal{T}}

\def\cS{\mathcal{S}}
\newcommand{\be}{\begin{equation}}
\newcommand{\ee}{\end{equation}}
\newcommand{\bea}{\begin{eqnarray}}
\newcommand{\eea}{\end{eqnarray}}

\begin{document}

\title{Modular invariant partition functions for non-compact $G/Ad(H)$ models}

\author{Jonas Bj\"ornsson$^1$~and~Jens Fjelstad$^2$}

\date{}

\maketitle
\void{
\vspace*{-40ex}
\hspace*{75ex}
{\small\vbox{\hbox{DAMTP-2010-130}}}
\vspace{35ex}}

\thispagestyle{myheadings}

\begin{center}
{\noindent $^1$Department of Applied Mathematics and
Theoretical Physics,\\ Wilberforce Road, Cambridge CB3 0WA, UK\\
{\tt j.bjornsson@damtp.cam.ac.uk}\\[2mm]
$^2$Department of Physics, Nanjing University,\\ 22 Hankou Road, Nanjing, 210098 China\\
{\tt jens.fjelstad@gmail.com}
}
\end{center}
\vspace{5mm}
\begin{abstract}
We propose a spectrum for a class of gauged non-compact $G/Ad(H)$ WZNW models, including spectrally flowed images of highest, lowest, and mixed extremal weight modules. These are combined into blocks whose characters, due to the Lorentzian signature of the target space, are divergent and treated as formal expressions in need of regularisation. Assuming that this is possible, we show that these extended characters transform linearly under modular transformations, and can be used to write down modular invariant partition functions.
\end{abstract}

\section{Introduction}
\label{sec:intro}

String propagation on curved non-compact manifolds is an interesting but difficult subject. Interesting as there are theories of this sort, which are dual to theories without gravity, one example being AdS/CFT. One complication is the curvature of space-time, and another that these target spaces usually come with non-zero $RR$-fluxes. This makes a covariant approach difficult.

There is, however, one class of theories with non-trivial target space geometry, but where an exact treatment is still viable. These are the gauged WZNW models \cite{Bardakci:1970nb,Goddard:1984vk}. The spacetime interpretations of these models are nevertheless complicated, and in all but a few exceptional cases still unclear. A subclass of non-compact gauged WZNW models have been elaborated on in \cite{Bjornsson:2007ha,Bjornsson:2008fq,Bjornsson:2009dp,Bjornsson:2010ja} where in particular, using a BRST construction following \cite{Karabali:1989dk}, sufficient conditions for a unitary string state space were determined. One may view this class of models as generalisations of the $AdS_3$ string, as described by a WZNW model based on the algebra $\mf{sl}(2,\R)$ \cite{Hwang:1990aq,Henningson:1991jc,Maldacena:2000hw}, and of the Nappi--Witten string \cite{Nappi:1993ie,Kiritsis:1993jk,Chen:2009wh}. Relevant for this class of models are groups $G$ admitting non-compact hermitian symmetric spaces of the form $G/H'$, where $H'$ is the maximal compact subgroup of $G$. It is known that $H'$ contains a one-dimensional center, and can locally be written as $H'=H\times U(1)$. The models in question are then obtained by gauging the subgroup $H$ in the $G$ WZNW model. As such these models may alternatively be thought of as conformal fix points of sigma models with semiclassical target spaces $G/Ad(H)$. The spaces $G/Ad(H)$ will then have a one compact direction in addition to the non-compact ones, and this direction will play the role of time.

In this paper we continue the investigation of the aforementioned non-compact gauged WZNW models with the purpose of constructing consistent string models. In this direction we propose a set of coset modules that close under modular transformations. What prevents us from showing this property strictly is that the naive combinations of characters of the proposed set of modules are divergent, and our proof of modular invariance is therefore on a formal level. This is not unique for our class of models, however, since also the $SL(2,\R)$ WZNW model on a finite cover has similar problems \cite{Henningson:1991jc}. The corresponding model on the universal cover also suffers from a divergent torus partition function, but this is interpreted as an analytic continuation of the well defined partition function of the $SL(2,\C)/SU(2)$ WZNW model \cite{Maldacena:2000hw,Gawedzki:1991yu}.

The structure of this paper is as follows.
In section \ref{sec:NgWZNW} we review briefly the BRST construction of non-compact gauged WZNW models from \cite{Bjornsson:2007ha}, which is based on previous work in \cite{Karabali:1989dk}. This is followed in section \ref{sec:SF} by a discussion of spectral flow automorphisms of a general affine Lie algebra, and the corresponding action on sets of modules of that algebra. The motivation for including modules generated by spectral flow is the unnatural bound on the string excitation level posed by unitarity analogously to the $SL(2,\R)$ model. Modules generated by spectral flow from highest, lowest, and mixed extremal weight modules, related by the finite Weyl group, are in section \ref{sec:Properties} grouped together in blocks, and the corresponding extended characters determined (this procedure generalises methods used in the $SL(2,\R)$ model). The construction of the extended characters is formal since different parts have different domains of convergence. Treating the extended characters as formal expressions we investigate their modular properties. We proceed in section \ref{sec:ModularPart}  by writing down partition functions that are, at least formally, modular invariant.

\section{Non-compact gauged WZNW models}
\label{sec:NgWZNW}

We begin with a short review of the BRST construction of a class of non-compact gauged WZNW models proposed in \cite{Bjornsson:2007ha}, which is based on \cite{Karabali:1989dk}. 


Let us introduce some notation and definitions, which mainly follow the conventions and notation used in \cite{Fuchs:1997jv}. Denote by $\mf{g}$ and $\mf{h}'$ the Lie algebras corresponding to a non-compact group $G$ and its maximal compact subgroup $H'$, which admits a Hermitian symmetric space of the form $G/H'$. Let $\mf{g}^{\C}$ and ${\mf{h}'}^{\C}$ denote the corresponding complex Lie algebras. We will always take the rank $r_{\mf{g}}$ of $\mf{g}$ to be greater than one and $\mf{g}$ to be simple. One knows that $\mf{h}'$ has a one-dimensional center, and therefore can be written as $\mf{h}\oplus \mf{u}(1)$. We choose $\mf{h}^{\C}$ such that it is a regular embedding in $\mf{g}^{\C}$, i.e.\ using the Cartan-Weyl basis the Cartan elements of $\mf{h}^{\C}$, as well as generators corresponding to positive/negative roots, are all elements of the corresponding basis of $\mf{g}^{\C}$. 

Denote by $\Delta$ all roots, $\Delta_{+/-}$ the positive/negative roots, $\Delta^c$ the compact roots, $\Delta^c_+=\Delta^c\cap\Delta_+$ the compact positive roots, $\Delta^n$ the non-compact roots and $\Delta^n_+$ the positive non-compact roots. We normalise the long roots to have length $\sqrt 2$. Let $\alpha\in \Delta$ and define the coroot by $\alpha^\vee=2\left(\alpha,\alpha\right)^{-1}\alpha$. Let $\alpha^{(i)}\in \Delta_+$ denote a chosen set of simple roots. Denote by $P$ the weight lattice and by $L^{\vee}$ the coroot lattice. When we need to distinguish between different root systems, we denote by $\Delta_{\mf g}$ and $\Delta_{\mf h}$ the roots in $\mf g^{\C}$ and ${\mf{h}}^{\C}$ respectively. Furthermore, we use capital letters, $A,B,\ldots$, and lowercase letters, $a,b,\ldots$, to denote elements in the algebra $\mf{g}^{\C}$ and $\mf{h}^{\C}$ respectively. We choose the set of simple roots of $\mf{g}$ such that the highest root is non-compact, and furthermore restrict the basis such that if $\alpha\in\Delta_{\mf g}^c$ then the first component is zero. With this choice we have a natural identification of $\Delta^c_{\mf{g}}$ with $\Delta_{\mf{h}}$. The dual Coxeter number  of $\mf{g}^{\C}$ is denoted $g_{\mf{g}}^\vee$. It is well known that one can choose simple roots such that there is a unique non-compact simple root, and if $\alpha\in\Delta^n_+$ then the coefficient of the non-compact simple root is always one in a simple root decomposition of $\alpha$. Dynkin diagrams and relations between positive non-compact roots of the Lie algebras are presented in the appendix of \cite{Jakobsen:1983} and, with our notation, in \cite{Bjornsson:2007ha}. For the embedding of $\mf{h}$ in $\mf{g}$, one defines the Dynkin index
\eqnb
I_{\mf{h}\subset\mf{g}}
      &=&
          \frac{\left(\theta(\mf{g}),\theta(\mf{g})\right)}{\left(\theta(\mf{h}),\theta(\mf{h})\right)}\,,
\eqne
where, for regular embeddings, $\theta(\mf{h})$ is the highest root of $\mf{h}^{\C}$ in $\mf{g}^{\C}$. 
We denote the untwisted affine Lie algebras corresponding to $\mf{g}^\C$ and $\mf{h}^\C$ by $\hat{\mf{g}}$ and $\hat{\mf{h}}$ respectively. When there is no risk of confusion between the finite dimensional Lie algebras and their affine counterparts we may sometimes simplify the notation by dropping the hats.
We will consider a coset CFT where the conformal anomaly takes the form
\eqnb
c_{\mathrm{tot}}
      &=&
          \frac{k d_{\mf{g}}}{k + g^{\vee}_{\mf{g}}}
          -\frac{\kappa d_{\mf{h}}}{\kappa + g^{\vee}_{\mf{h}}}\,,
\label{conformalanomaly}
\eqne
where $d_{\mf{g}}$ and $d_{\mf{h}}$ denote the dimensions of $\mf{g}^{\C}$ and $\mf{h}^{\C}$, respectively. In (\ref{conformalanomaly}) $\kappa$ is defined as
\eqnb
\kappa&\equiv&I_{\mf{h}\subset\mf{g}}k\,,
\eqne
for an integer level $k$ such that $k+g_\mf{g}^\vee<0$.

We will in this paper use radial ordering,
\eqnb
\mathcal{R}(A(z)B(w))
      &=&
          \left\{
          \begin{array}{rr}
              A(z)B(w) & \left|z\right| > \left|w\right| \\
              \left(-\right)^{\epsilon_{A}\epsilon_{B}}B(w)A(z) & \left|z\right| < \left|w\right|
          \end{array}
          \right.\,,
\eqne
where $\epsilon$ denote the Grassman parity of the operator. As usual we suppress the symbol $\mathcal{R}$, and the radially ordered product satisfies
\eqnb
A(z)B(w)
	& = & :AB(w): + \overbracket{A(z)B(w)} +\, \mathcal{O}(z-w)\;\; \sim\;\; \overbracket{A(z)B(w)}\,,
\eqne
where $:\;\;:$ denotes normal ordering, $\overbracket{\phantom{123}}$ the contraction, and $\sim$ denotes equality up to regular terms. 
From this one obtains
\eqnb
:AB(z):
	&=&	
		\frac{1}{2\pi i} \oint_{z} dw (w-z)^{-1}A(w)B(z)\,.
\label{normaltwofields}
\eqne

The OPE and commutator of Laurent modes of operators are related as 
\eqnb
\left[A_m,B_n\right]
      &=&
          \frac{1}{\left(2\pi i\right)^2}\oint_0 dw w^{\Delta_B+n-1}\oint_w dz z^{\Delta_A+m-1} A(z)B(w)\,,
\eqne
where $\Delta_A$ and $\Delta_B$ are the conformal weights of the operators $A$ and $B$ respectively.

Denote by $g(z,\bar{z})$ the map from the (Euclidean) worldsheet to $G$, then the holomorphic and anti-holomorphic currents of the WZNW model on $G$ are
\eqnb
J_L(z) &=& k\, g^{-1}\partial g \no
J_R(\bar{z}) &=& -k\, \bar{\partial} g g^{-1}\,.
\eqne
We will mainly consider the holomorphic part in this paper as the anti-holomorphic part works in the same way, and therefore we suppress the subscript $L$. The currents take values in the Lie algebra $\mf{g}^\C$ and are expressed in components by using a basis of the algebra, $J^A(z)=\kappa\left(t^A,J(z)\right)$ where $\kappa(\cdot,\cdot)$ is the killing form. The OPE between two currents takes the form
\eqnb
J^A(z)J^B(w) &\sim& k \kappa^{AB} \left(z-w\right)^{-2} + {f^{AB}}_C J^C(w)(w-z)^{-1}\,.
\eqne
In the Cartan-Weyl basis, the structure constants and killing form are
\eqnb
{f^{i\alpha}}_\beta
      =
          \alpha^{i}\delta_{\alpha,\beta}\, ,
      &\phantom{123}&
{f^{\alpha\beta}}_i
      =
          \alpha^\vee_i\delta_{\alpha+\beta,0}\, ,
      \no
{f^{\alpha\beta}}_{\gamma}
      =
          e_{\alpha,\beta}\delta_{\alpha+\beta,\gamma}\, ,
      \no
\kappa^{ij}
      =
          G_{\mf{g}}^{ij}\, ,
      &\phantom{123}&
\kappa^{\alpha\beta}
      =
          \frac{2}{\left(\alpha,\alpha\right)}\delta_{\alpha+\beta,0}\, ,
\eqne
where $e_{\alpha,\beta}$ are only non-zero if $\alpha+\beta$ is a root and $G_{\mf{g}}^{ij}\equiv({\alpha^{(i)}}^\vee,{\alpha^{(j)}}^\vee)$ is the inverse of the quadratic form $G^{\mf{g}}_{ij}$ of $\mf{g}^{\C}$.
 
We will use the BRST approach to construct the coset \cite{Karabali:1989dk} as this was the approach where unitarity was achieved \cite{Bjornsson:2007ha}. For compact algebras the BRST and GKO constructions have been shown to have equivalent spectrum \cite{Hwang:1993nc}. Apart from adding a ghost sector to the $G$ WZNW model at level $k$, one needs to add an auxiliary sector which is described by a $H$ WZNW model of level $\tilde{\kappa} = -\kappa - 2g_{\mf{h}}$. We will abbreviate $\hat{\tilde{\mf{h}}} = \mf{h}_{\tilde{\kappa}}$ and denote the corresponding currents by $\tilde{J}_L^a(z)$ and $\tilde{J}_R^a(\bar{z})$. Using this sector, the constraints imposed on the physical states are changed from $J^+_{\mf{h}}\left|\Phi\right> = 0$ to $\left(J^+_{\mf{h}}+J^+_{\tilde{\mf{h}}}\right)\left|\Phi\right> = 0$, where $J^+_{\bullet}$ denotes an arbitrary annihilation operator in the chosen module. 

From the current-current OPE one can obtain a BRST current for the coset constraints
\eqnb
j^1_{BRST}(z)
      &=&
          :c_a\left(J^a + \tilde{J}^a\right): - \frac{1}{2}{f^{ab}}_c:c_ac_bb^c: \, ,
\label{BRST.coset}
\eqne
where $c_a$ are fermionic ghosts with momenta $b^a$ satisfying the OPE
\eqnb
c_a(z)b^b(w)
   &\sim&
      \delta_{a}^{b}(z-w)^{-1}\, .
\eqne
The OPE of the current (\ref{BRST.coset}) with itself has only regular terms as $\kappa + \tilde{\kappa} + 2g^{\vee}_{\mf{h}'}= 0$, and we therefore have a nilpotent BRST operator. The point of \cite{Bjornsson:2007ha} was to find non-compact coset models with unitary {\em string} spectra, and this was indeed achieved by combining the coset constraints with the Virasoro constraints in a BRST current:
\eqnb
j_{BRST}(z)
	&=&
		:c_a\left(J^a+\tilde{J}^a\right): + :cT: - \frac{1}{2}{f^{ab}}_c:c_ac_bb^c :
	\no 
		&-& :\partial c cb: + :c\partial b^a c_a: + :\partial c b^ac_a: \, .
\label{BRST-current}
\eqne
Here the energy momentum tensor $T(z)$ is given by
\eqnb
T(z)
   &=&
      \frac{1}{2\left(k + g^{\vee}_{\mf{g}}\right)}\kappa_{AB} :J^AJ^B(z): + \frac{1}{2\left(\tilde{\kappa} + g^{\vee}_{\mf{h}}\right)}\kappa_{ab} :\tilde{J}^a\tilde{J}^b(z): + T'(z)\, ,
\eqne
where $T'(z)$ is the energy momentum tensor of a unitary conformal field theory with conformal anomaly $c'$ added to give a critical central charge (i.e.\ with value 26).
Note that the Virasoro ghosts are coupled to the coset ghosts, so in the present formalism we are really constructing a coset string model in one step. It is possible, however, to first gauge the subgroup $H$ to get a coset CFT, and only later apply the Virasoro constraints.

The Laurent expansion of the currents and the energy momentum tensor are
\eqnb
J^i(z)
	&=&
		\sum_{n\in \Z} H^i_n z^{-n-1}
	\no
J^\alpha(z)
	&=&
		\sum_{n\in \Z} E^\alpha_n z^{-n-1}
	\no
T(z)
	&=&
		\sum_{n\in \Z} L_n z^{-n-2}\,.
\eqne
In $\hat{\mf{g}}$ there exists a set of special elements defined by the current
\eqnb
H(z)
	&\equiv&
		\Lambda_{(1),i},J^i(z)
	\no
	&=&
		\left(\Lambda_{(1)},J(z)\right)\,,
\eqne
which has only regular terms in the OPE with any current corresponding to the subalgebra $\mf{h}$. An excitation by a creation operator of this current gives a state with negative norm, i.e. the direction on $G$ specified by the zero-mode of $H(z)$ is timelike.

From the BRST current one can define useful charges by commuting the BRST charge with zero modes of the $b$ ghosts of the Cartan subalgebra and the reparametrisation (Virasoro) ghost:
\eqnb
L^{\mathrm{tot}}_0
	&=&
		[Q,b_0]
	\no
	&=&
		L^{\mf{g}}_0 + L^{\tilde{\mf{h}}}_0 + L'_{0} + L^{\mathrm{Vir}\; gh}_0 + L^{\mathrm{\mf{g}/\mf{h}}\; gh}_0 - 1
	\\
H^{i;\mathrm{tot}}_0
	&=&
		[Q,b_0^i]
	\no
	&=&
		H^i_0 + \tilde{H}^i_0 + H^{i; gh}_0\, ,
\eqne
where
\eqnb
L^{\mf{g}}_0
	&=&
		\frac{1}{2(k+g_{\mf{g}}^{\vee})}\sum_{n\in\Z}\left(\sum_{i=1}^{r_{\mf{g}}} G^\mf{g}_{ij}:H^i_{-n}H^j_{n}: + \sum_{\alpha\in\Delta_{\mf{g}}}\frac{(\alpha,\alpha)}{2}:E^{\alpha}_{-n}E^{-\alpha}_{n}:\right)
	\no
L^{\tilde{\mf{h}}}_0
	&=&
		\frac{1}{2(\kappa+g_{\mf{h}}^{\vee})}\sum_{n\in\Z}\left(\sum_{i=2}^{r_{\mf{g}}} G^\mf{h}_{ij}:\tilde{H}^i_{-n}\tilde{H}^j_{n}: + \sum_{\alpha\in\Delta_{\mf{h}}}\frac{(\alpha,\alpha)}{2}:\tilde{E}^{\alpha}_{-n}\tilde{E}^{-\alpha}_{n}:\right)
	\no
L^{\mathrm{\mf{g}/\mf{h}}\; gh}_0
	&=&
		\sum_{n\in\Z}n:b^a_{-n}c_{n,a}:
	\no
L^{\mathrm{Vir}\; gh}_0
	&=&
		\sum_{n\in \Z }n:b_{-n} c_{n}:
	\no
H^{i; gh}_0
	&=&
		\sum_{n,\alpha\in\Delta_\mf{h}}\alpha^i:b^\alpha_{-n}c_{n,\alpha}:\, .
\eqne

In determining a modular invariant spectrum we will need to impose some restrictions on the modules of $\mf{g}$ and $\tilde{\mf{h}}$. We will here describe a certain subset of these modules that will serve as a starting point, and additional modules will be generated further on. Consider first the $\mf{g}$ part of the module space. The modules that we start out with are constructed by imposing the highest weight conditions
\eqnb
E_n^{\alpha}\left|\mu;k,0\right>
	&=&
		0\phantom{\mu^i\left|\mu;k,0\right>123} \alpha \in \Delta ,\;\; n > 0
	\no
E_0^{\alpha}\left|\mu;k,0\right>
	&=&
		0\phantom{\mu^i\left|\mu;k,0\right>123} \alpha \in \Delta_+
	\no
H^i_n\left|\mu;k,0\right>
	&=&
		0\phantom{\mu^i\left|\mu;k,0\right>123} n>0
	\no
H_0^i\left|\mu;k,0\right>
	&=&
		\mu^i\left|\mu;k,0\right>\, ,
\eqne
and the module is generated by acting with the creation operators modulo null states. We will for the $\mf{g}$ module consider integral anti-dominant highest weights satisfying
\eqnb
\mu^i &\leq& -2 \phantom{12345} i=1,\ldots, r_\mf{g}\no
k + 2 &\leq& \left(\mu,\theta\right)\,.
\eqne
Note that this requires $k+g_{\mf{g}}^{\vee}< 0$. We will in addition require that $k\in \Z$. This set of highest weights we denote by $P^{k,\mf{g}}_{-}$. A unitary string spectrum poses less restrictions \cite{Bjornsson:2007ha,Bjornsson:2009dp}, however this set will prove to be enough in order to find modular invariant partition functions. The highest weight modules freely constructed from these weights do not have any null states, and the corresponding irreducible modules are therefore Verma modules. Denote the module with highest weight $\mu$ at level $k$ by $\mathcal{F}^{k,\mf{g}}_{\mu}$. In order to consider modules for the relevant real form we impose the following adjointness conditions
\eqnb
\left(E^{\alpha}_n\right)^{\dagger}
	&=&
		E^{-\alpha}_{-n} \phantom{-E^{-\alpha}_{-n}} \alpha\in \Delta^{c} 
	\no
\left(H^{i}_n\right)^{\dagger}
	&=&
		H^i_{-n} 
	\no
\left(E^{\alpha}_n\right)^{\dagger}
	&=&
		-E^{-\alpha}_{-n} \phantom{E^{-\alpha}_{-n}} \alpha\in \Delta^{n}\, .
\eqne
The modules we choose for $\mf{h}$ are constructed analogously, with the difference that we then consider dominant integral highest weights,
\eqnb
P^{\tilde{\kappa},\mf{h}}_{+}
	&\equiv&
		\{\tilde{\mu} \in P^\mf{h}: \tilde{\mu}^i \geq 0\;\;i=2,\ldots, r_\mf{g}\,,\,\left(\tilde{\mu},\theta_{\mf{h}}\right)\leq \tilde{\kappa} \}\,.
\eqne
Denote the module with highest weight $\tilde{\mu}$ at level $\tilde{\kappa}$ by $\mathcal{F}^{\tilde{\kappa},\mf{h}}_{\tilde{\mu}}$.

The coset ghost sector of the theory is constructed from a ghost vacuum using the operators $c^{\alpha}_n\equiv c_{n,-\alpha}$, satisfying
\eqnb
b_n^i\left| 0\right>_{b,c} = 0\,, \phantom{12} n\geq 0\,,
&\phantom{123}&
b_n^\alpha\left| 0\right>_{b,c} = c_n^\alpha\left| 0\right>_{b,c} = 0\,, \phantom{12} n > 0 \;\;\alpha\in\Delta\,,
\no
c_{n,i}\left| 0\right>_{b,c} = 0\,, \phantom{12} n > 0 \,,
&\phantom{123}&
b_0^\alpha\left| 0\right>_{b,c} = c_0^\alpha\left| 0\right>_{b,c} = 0\,, \phantom{12} \alpha\in\Delta_+\,.
\label{ghostvac}
\eqne
The vacuum state $\left|0\right>_{b,c}$ is therefore a ``down-state''. Denote this module by $\mathcal{G}^{\mf{g}/\mf{h}}$. Using the conditions on the highest weight, it follows that
\eqnb
H^{i; gh}\left|0\right>_{b,c}
	&=&
		2\rho^i\left|0\right>_{b,c} \;\;=\;\; 2\left|0\right>_{b,c} \phantom{123} i=2,\ldots,r_\mf{g}
	\no
L_0^{\mathrm{Vir}\,gh}\left|0\right>_{b,c}
	&=&
		2g^{\vee}\left|0\right>_{b,c}\,.
\eqne
The reparametrisation ghost sector is determined in the standard way, and we denote the corresponding module by $\mathcal{G}^{\mathrm{Vir}}$. The unitary CFT may contain a number of sectors, which we label by $\lambda$, and the corresponding modules are denoted $\mathcal{F}_{\lambda}$. A single sector of the full theory, including irreducible modules of $\mf{g}$, $\tilde{\mf{h}}$, ghosts and the unitary CFT, will be denoted by $\mathcal{F}_{\mu,\tilde{\mu},\lambda}^k$.

Imposing the condition that the physical states in $\mathcal{F}_{\mu,\tilde{\mu},\lambda}^k$ are BRST invariant modulo trivial states,
\eqnb
Q\left|\Phi\right> &=& 0 \no
\left|\Phi\right> &\sim& \left|\Phi\right> + Q\left|\chi\right>\, ,
\eqne
yields $2^{r_{\mf{g}}}$ copies of the physical states with different ghost number. The surplus of states can be projected out by imposing additional conditions
\eqnb
b_0\left|\Phi\right> = b^i_0\left|\Phi\right> = 0\,.
\eqne
This results in all physical states having the same ghost number. Denote the space spanned by these states by $\mathcal{H}_{\mu,\tilde{\mu},\lambda}^k$. Observe that these additional conditions also imply matching conditions between different weights of the modules
\eqnb
H^{i;\mathrm{tot}}_0\left|\Phi\right> &=& [Q,b^i_0]\left|\Phi\right>\;\; =\;\; 0 \no
L^{\mathrm{tot}}_0\left|\Phi\right> &=& [Q,b_0]\left|\Phi\right> \;\; =\;\; 0 \, .
\eqne
This will be achieved in the end by integrating over certain parameters of the theory.

We can define two traces over all states in $\mathcal{F}^{k}_{\mu,\tilde{\mu},\lambda}$ with a non-zero contribution only from the physical subspace $\mathcal{H}^{k}_{\mu,\tilde{\mu},\lambda}$. These traces are the character and signature function respectively,
\eqnb
\chi_{\mu,\tilde{\mu},\lambda}^k(\Phi;\tau,t_{\mf{g}/\mf{h}})
	&\equiv&
		\Tr_{\mathcal{F}^{k}_{\mu,\tilde{\mu},\lambda}}[(-1)^{\Delta N_{gh}}e^{2\pi i \tau L^{\mathrm{tot}}_0}e^{2\pi i \left(\Phi,H_0^{tot}\right)}e^{2\pi i kt_{\mf{g}/\mf{h}}}]
	\label{fullchar}
	\\
\Sigma_{\mu,\tilde{\mu},\lambda}^k(\Phi;\tau,t_{\mf{g}/\mf{h}})
	&\equiv&
		\Tr'_{\mathcal{F}^{k}_{\mu,\tilde{\mu},\lambda}}[(-1)^{\Delta N_{gh}}e^{2\pi i \tau L^{\mathrm{tot}}_0}e^{2\pi i \left(\Phi,H_0^{tot}\right)}e^{2\pi i kt_{\mf{g}/\mf{h}}}]\,,
\eqne
where $\Tr'_{\mathcal{F}^{k}_{\mu,\tilde{\mu},\lambda}}$ is the trace with signature of the state and $\Delta N_{gh}$ is the ghost number relative to the ground state. In the variable $\Phi= (\theta,\phi)$, $\theta_i$ multiply the elements in the Cartan sub-algebra which are common to $\mf{h}$ while $\phi$ multiplies the element $H_0 = (\Lambda_{(1)},H_0)$ which commutes with all elements in $\mf{h}$. To get the true character and signature function for $\mathcal{H}^{k}_{\mu,\tilde{\mu},\lambda}$, however, one must integrate over $\theta$ and $\tau$ to project to the correct subspace. In \cite{Bjornsson:2007ha,Bjornsson:2009dp} theorems concerning unitarity of $\mathcal{H}^{k}_{\mu,\tilde{\mu},\lambda}$ were proven. The sufficient conditions state in particular that $\mu$ is integer for the parts contained in the subalgebra $\mf{h}$. The proof compares the character and signature function.

As the trace is over all states in the theory, the character and the signature function can be decomposed into the different sectors
\eqnb
\chi_{\mu,\tilde{\mu},\lambda}^k(\Phi;\tau,t_{\mf{g}/\mf{h}})
	&=&
		e^{2\pi i \tau \frac{c_\mf{g} + c_{\tilde{\mf{h}}} + 2g^{\vee}_\mf{h}+c'-26}{24}}\chi_{\mu}^{k,\mf{g}}(\Phi;\tau,t_{\mf{g}/\mf{h}},t_\mf{h})\chi_{\tilde{\mu}}^{\tilde{\kappa},\tilde{\mf{h}}}(\theta;\tau,t_\mf{h})
	\no
	&\times&
		\chi^{\mf{g}/\mf{h}\;gh}(\theta;\tau,t_\mf{h})\chi^{\mathrm{Vir}\;gh}(\tau)\chi^{\mathrm{CFT}}_{\lambda}(\tau)\, ,
	\label{fullchardecom}
\eqne
here the overall factor is equal to one, as we have assumed that the total conformal charge is $26$. The characters in the decomposition are
\eqnb
\chi_{\mu}^{k,\mf{g}}(\Phi;\tau,t_{\mf{g}/\mf{h}},t_\mf{h})
	&\equiv&
		\Tr_{\mathcal{F}_{\mu}^{k,\mf{g}}}[e^{2\pi i \tau \left(L^\mf{g}_0-\frac{c_\mf{g}}{24}\right)}e^{2\pi i \left(\Phi,H_0\right)}e^{2\pi i kt_{\mf{g}/\mf{h}}+2\pi i \kappa t_\mf{h}}]
	\label{charG}
	\\
\chi_{\tilde{\mu}}^{\tilde{\kappa},\tilde{\mf{h}}}(\theta;\tau,t_\mf{h})
	&\equiv&
		\Tr_{\mathcal{F}_{\tilde{\mu}}^{\tilde{\kappa},\mf{h}}}[e^{2\pi i \tau \left(L^{\tilde{\mf{h}}}_0-\frac{c_{\tilde{\mf{h}}}}{24}\right)}e^{2\pi i \left(\theta,\tilde{H}_0\right)}e^{2\pi i \tilde{\kappa} t_\mf{h}}]
	\label{charH}
	\\
\chi^{\mf{g}/\mf{h}\;gh}(\theta;\tau,t_\mf{h})
	&\equiv&
		\Tr_{\mathcal{G}^{\mf{g}/\mf{h}}}[(-1)^{\Delta N_{gh}}e^{2\pi i \tau \left(L^{\mf{g}/\mf{h}\; gh}_0-\frac{2g^{\vee}_\mf{h}}{24}\right)}e^{2\pi i \left(\theta,H^{gh}_0\right)}e^{4\pi i g^{\vee}_\mf{h} t_\mf{h}}]
	\label{charG/H}
	\\
\chi^{\mathrm{Vir}\;gh}(\tau)
	&\equiv&
		\Tr_{\mathcal{G}^{\mathrm{Vir}}}[(-1)^{\Delta N_{gh}}e^{2\pi i \tau \left(L^{\mf{g}/\mf{h}\; gh}_0-\frac{1}{12}\right)}]
	\label{charVIR}
	\\
\chi^{\mathrm{CFT}}_{\lambda}(\tau)
	&=&
		\Tr_{\mathcal{F}_{\lambda}}[e^{2\pi i \tau \left(L^{\mathrm{CFT}}_0-\frac{c'}{24}\right)}]
		\,.
	\label{charUCFT}
\eqne
In the two equations above we have introduced another variable, $t_\mf{h}$, which cancels in the product because $\kappa + \tilde{\kappa} + 2g^{\vee}_\mf{h} = 0$. The introduction of this variable serves to make the modular properties of the functions more transparent.

From here on we will manly discuss $\chi_{\mu}^{k,\mf{g}}$. The properties of the other sectors are well known, and we will in most cases only state the results of computations.

\section{Spectral Flow}
\label{sec:SF}

Due to the $\mf{g}$-sector there is no obvious reason why the combined characters written in the last section should behave nicely under modular transformations, even less so considering the need to integrate over the parameter $\theta$. Therefore it seems unlikely that it is possible to construct a well-defined theory with such a spectrum. One is faced with a similar problem in the $SL(2,\R)$ WZNW model if one tries to define a theory with spectrum from the discrete series satisfying the unitarity bound but in that case we know a way out \cite{Henningson:1991jc,Maldacena:2000hw}, namely to extend the spectrum with an infinite number of additional sectors generated by spectral flow automorphisms of the affine algebra\footnote{The exact procedure differs on a finite cover \cite{Henningson:1991jc} compared to the infinite cover \cite{Maldacena:2000hw} of $SL(2,\R)$, but in both cases one extends by modules generated by spectral flow. In addition it is known, at least on the infinite cover, that the fusion rules require also adding the principal continuous series to the spectrum.}. An optimistic guess would then be that a similar extension is possible in the class of coset models considered here, and as we will see this does indeed work. Let us first define an infinite set of automorphisms of the affine Lie algebra $\hat{\mf{g}}$ that we will refer to as spectral flow.

Let $w^L_\mf{g}$ and $w^R_\mf{g}$ be two arbitrary weights of the horizontal subalgebra of $\hat{\mf{g}}$, then spectral flow with these parameters should have the following effect on a map $g(z,\bar{z})$:
\eqnb
g(z,\bar{z}) \mapsto \bar{z}^{\left(w_{\mf{g}}^R, H_0\right)} g(z,\bar{z})  {z}^{\left(w_{\mf{g}}^L, H_0\right)}\,.
\eqne
We restrict our discussion to the holomorphic part, the anti-holomorphic works out in the same way, and we skip the labels $L$ and $\mf{g}$ on the parameter $w_{\mf{g}}^{L}$. The current $J(z)$ transforms as
\eqnb
J(z) &\mapsto& z^{-(w,H_0)} J(z) z^{(w,H_0)} + k\frac{(w,H_0)}{z}\,.
\eqne
In the Cartan-Weyl basis of the Lie algebra, the effect of spectral flow is
\eqnb
J^i(z)
	&\mapsto&
		J^i(z) + k\frac{G_{\mf{g}}^{ij}w_j}{z} \no
J^{\alpha}(z)
	&\mapsto&
		z^{(w,\alpha)}J^{\alpha}(z)\,.
\eqne
Demanding that the current is still holomorphic after the action we see that $w$ is restricted to the coroot lattice, but there are no further restrictions.
The Laurent modes of the fields transform as
\eqnb
H^i_n
	&\mapsto&
		H^i_n + kG_{\mf{g}}^{ij}w\delta_{n,0} \label{CartanSpFl}\\
E^{\alpha}_n
	&\mapsto&
		E_{n+(w,\alpha)}\,.
\eqne
It is now straightforward to show that the spectral flow is an automorphism of the algebra when $(w,\alpha)\in\Z$. More precisely, composing the action above with parameters $w$ and $w'$ we get an action of $w+w'$. It is thus immediate that we are considering an infinite subgroup of $\mathrm{Aut}(\hat{\mf{g}})$ that is isomorphic to the coroot lattice $L^\vee_\mf{g}$.

Any automorphism $\varphi$ of an algebra $A$ will give an action on the category of $A$-modules. Consider a module $M=(V,\rho)$, where $V$ is the underlying vector space and $\rho:A\rightarrow \mathrm{End}(V)$ is the representation morphism. The action of $\varphi$ on this module gives the module $M^\varphi=(V,\rho\circ\varphi)$. Therefore we can generate new modules of $\hat{\mf{g}}$ using the spectral flow automorphisms. The characters of these modules will be important to us, so we proceed by determining the action of $L^\mf{g}_0$ on the modules generated by spectral flow (the action of $H_0^i$ is read off from (\ref{CartanSpFl})). Consider the energy momentum tensor
\eqnb
T_{\mf{g}}(z) &=& \frac{1}{2\left(k+g^{\vee}\right)}\kappa_{AB}:J^AJ^B(z):\,.
\eqne
We can determine from the currents how $T_{\mf{g}}(z)$ transforms using (\ref{normaltwofields}). Consider first $(A,B) = (i,j)$,
\eqnb
:J^iJ^j(z):
	&\mapsto&
		\frac{1}{2\pi i} \oint_z \frac{dz'}{z'-z} \left[\left(J^i(z') + k\frac{G_{\mf{g}}^{ik}w_k}{z'}\right)\left(J^j(z) + k\frac{G_{\mf{g}}^{jl}w_l}{z}\right)\right]
	\no
	&=&
		\frac{1}{2\pi i} \oint_z \frac{dz'}{z'-z} \left(J^i(z')J^j(z)\right)
	+
		\frac{k}{2\pi i} \oint_z \frac{dz'}{z'-z} \left(\frac{J^i(z')}{z}\right)G_{\mf{g}}^{jl}w_l
	\no
	&+&
		\frac{k}{2\pi i} \oint_z \frac{dz'}{z'-z} \left(\frac{J^j(z)}{z'}\right)G_{\mf{g}}^{ik}w_k
	+
		\frac{k^2}{2\pi i} \oint_z \frac{dz'}{z'-z} \left(\frac{1}{zz'}\right)G_{\mf{g}}^{ik}w_k G_{\mf{g}}^{jl}w_l
	\no
	&=&
		:J^iJ^j(z):
		+ k \frac{J^i(z)}{z} G_{\mf{g}}^{jk}w_k + k \frac{J^j(z)}{z} G_{\mf{g}}^{ik}w_k + \frac{k^2}{z^2}G_{\mf{g}}^{ik}G_{\mf{g}}^{jl}w_kw_l\,.
\eqne
For $(A,B) = (\alpha,-\alpha)$ one obtains
\eqnb
:J^{\alpha}J^{-\alpha}(z): 
	&\mapsto&
		\frac{1}{2\pi i} \oint_z \frac{dz'}{z'-z} \left( z'^{(w,\alpha)}J^{\alpha}(z')z^{(w,\alpha)}J^{-\alpha}(z) \right)
	\no
	&=&
	\sum_{m,n=-\infty}^{\infty} E_m^{\alpha}E_n^{-\alpha}\frac{1}{2\pi i} \oint_z \frac{dz'}{z'-z} \left[\left(\frac{z'}{z}\right)^{(w,\alpha)}z'^{-m-1} z^{-n-1}\right]\,.
\eqne
The integral is determined in the standard way by splitting it up into two different regions where different expansions are valid and one obtains
\eqnb
:J^{\alpha}J^{-\alpha}(z):
	&\mapsto&
		\sum_{n=-\infty}^{\infty}\left(\sum_{m=-\infty}^{(w,\alpha)-1}E^{\alpha}_mE^{-\alpha}_n + \sum_{m = (\alpha,w)}^{\infty}E^{-\alpha}_nE^{\alpha}_m\right)z^{-m-n-2}\,.
\eqne
This expression should be reordered to be compatible with the normal ordering of the module without spectral flow. Consider first the case $(w,\alpha)>0$, this gives
\eqnb
&&\sum_{m=-\infty}^{(w,\alpha)-1}E^{\alpha}_mE^{-\alpha}_n + \sum_{m = (\alpha,w)}^{\infty}E^{-\alpha}_nE^{\alpha}_m - \sum_{m=-\infty}^{-1}E^{\alpha}_mE^{-\alpha}_n - \sum_{m = 0}^{\infty}E^{-\alpha}_nE^{\alpha}_m
	\no
	&&=
		\sum_{m=0}^{(w,\alpha)-1}[E^{\alpha}_m,E^{-\alpha}_n]
	\no
	&&=
		\sum_{m=0}^{(w,\alpha)-1}\left(\alpha^{\vee}_i H_{m+n}^i + \frac{2}{(\alpha,\alpha)}mk\delta_{m+n,0}\right)
	\no
	&&=
		(w,\alpha)\alpha^{\vee}_i H^i_{m+n} + \frac{2}{(\alpha,\alpha)}\frac{1}{2}k(w,\alpha)\left((w,\alpha)-1\right)\delta_{n,0}\,.
\eqne
For the case $(w,\alpha)<0$ one obtains the same result, so we have shown
\eqnb
\hspace{-0.5cm}
:J^{\alpha}J^{-\alpha}(z):
	&\mapsto&
		:J^{\alpha}J^{-\alpha}(z): + (w,\alpha)\alpha^{\vee}_i \frac{J^i(z)}{z} + \frac{1}{2z^2}k(w,\alpha^{\vee})\left((w,\alpha)-1\right)\,.
\eqne
Summarising, spectral flow transforms the energy momentum tensor according to
\eqnb
T_{\mf{g}}(z)
	&\mapsto&
		T_{\mf{g}}(z) + \frac{1}{k+ g^{\vee}} \frac{\left(kG^{\mf{g}}_{ij} + \frac{1}{2}\sum_{\alpha\in\Delta_{\mf{g}}}\alpha_i\alpha_j\right) G_{\mf{g}}^{ik}w_k J^j(z)}{z} 
	\no
	&+&
		\frac{1}{2(k+ g^{\vee})}\left(k^2G^{\mf{g}}_{ij} + \frac{k}{2}\sum_{\alpha\in\Delta_{\mf{g}}}\alpha_i\alpha_j\right)G_{\mf{g}}^{ik}w_kG_{\mf{g}}^{jl}w_l\,.
\eqne
We can simplify this expression by using 
\eqnb
\sum_{\alpha\in\Delta_{\mf{g}}}\alpha_i\alpha_j = 2g^{\vee}G^{\mf{g}}_{ij}\,,
\eqne
to obtain
\eqnb
T_{\mf{g}}(z)
	&\mapsto&
		T_{\mf{g}}(z) + \frac{(w,J(z))}{z} + \frac{1}{2z^2}k(w,w)\,.
\eqne
Thus, the zero modes are transformed as
\eqnb
L^{\mf{g}}_0
	&\rightarrow&
		L^{\mf{g}}_0 + (w,H_0) + \frac{k}{2}(w,w)\label{L0sfl}
	\\
H^i_0
	&\rightarrow&
		H^i_0 + kG_{\mf{g}}^{ij}w_j\, .\label{H0sfl}
\eqne

Denote by $\cF_{\mu,w}^{k,\mf{g}}$ the module generated from $\cF_\mu^{k,\mf{g}}$ by the spectral flow $w$ (we identify $\cF_{\mu,0}^{k,\mf{g}}\equiv\cF_\mu^{k,\mf{g}}$). For non-zero $w$ the module $\cF^{k,\mf{g}}_{\mu,w}$ will no longer be highest weight, but it is still generated from a single vector (the highest weight vector in $\cF^{k,\mf{g}}_\mu$). It is clear that $\cF^{k,\mf{g}}_{\mu,w}$ is irreducible since $\cF^{k,\mf{g}}_\mu$ is irreducible. Furthermore, if $w\neq w'$ then $\cF^{k,\mf{g}}_{\mu,w}\ncong \cF^{k,\mf{g}}_{\mu,w'}$. Using the methods from \cite{Bjornsson:2007ha} it is straightforwardly shown that the modules generated by spectral flow also give unitary string states under the same conditions as for the $w=0$ sector.

The character of $\cF^{k,\mf{g}}_{\mu,w}$ is defined in the same way as that of $\cF^{k,\mf{g}}_{\mu}$ in (\ref{charG}) and one obtains:
\eqnb
\chi_{\mu,w}^{k,\mf{g}}(\Phi,\tau,t_{\mf{g}/\mf{h}},t_\mf{h}) 
	& \equiv &
		\Tr_{\cF^{k,\mf{g}}_{\mu,w}} \left[ e^{2\pi i\tau\left(L_0-\frac{c_\mf{g}}{24}\right)+2\pi i \left(\Phi,H_0\right) + 2\pi i k t_{\mf{g}/\mf{h}} + 2\pi i\kappa t_\mf{h}}\right]\\
	& = & 
		\Tr_{\cF^{k,\mf{g}}_\mu}\left[e^{2\pi i \tau \left(L_0 + (w,H_0) + \frac{k}{2}(w,w) - c_{\mf{g}}/24\right)+ 2\pi i (\Phi,H_0+kw) + 2\pi i k t_{\mf{g}/\mf{h}} + 2\pi i \kappa t_{\mf{h}}}\right]
	\no
	& = &
		e^{\pi i \tau \left(k(w,w) - c_{\mf{g}}/12 \right) + 2\pi i k (\Phi,w) + 2\pi i k t_{\mf{g}/\mf{h}} + 2\pi i \kappa t_{\mf{h}}} 
	\no
	& \times &
		\Tr_{\cF^{k,\mf{g}}_\mu}\left[e^{2\pi i \tau L_0 + 2\pi i (\Phi+w\tau,H_0)}\right] \,,
\eqne
where the second line follows immediately from (\ref{L0sfl}) and (\ref{H0sfl}).
The trace is convergent if $\tau$ lies in the upper half plane, and can be explicitly evaluated since $\cF^{k,\mf{g}}_\mu$ is a Verma module with the result
\eqnb
&&
\hspace{-0.5cm}
\chi_{\mu,w}^{k,\mf{g}}(\Phi;\tau,t_{\mf{g}/\mf{h}},t_{\mf{h}})
\no
&&=
		\frac{e^{2\pi i\tau \left(\frac{(\mu,\mu + 2\rho_\mf{g})}{2(k+g_\mf{g}^{\vee})} + (w,\mu) + \frac{k}{2}(w,w) - \frac{c_{\mf{g}}}{24}\right)+ 2\pi i (\Phi,\mu+kw) + 2\pi i k t_{\mf{g}/\mf{h}} + 2\pi i \kappa t_{\mf{h}}}}{\prod_{\alpha\in \Delta_{\mf{g},+}}\left(1-e^{-2\pi i(\alpha,w\tau + \alpha}\right)\prod_{n>0}(1-e^{2\pi i n \tau})^{r_{\mf{g}}}\prod_{\alpha\in \Delta_\mf{g}}\left(1-e^{2\pi i n\tau + 2\pi i(\alpha,w\tau + \alpha}\right) }
	\no
	&&=
		\frac{e^{2\pi i \left(\frac{(\mu,\mu + 2\rho_\mf{g})}{2(k+g_\mf{g}^{\vee})} + (w,\mu) + \frac{k}{2}(w,w) - \frac{c_{\mf{g}}}{24}\right)+ 2\pi i (\Phi,\mu+kw) + 2\pi i \left(k + g_\mf{g}^{\vee}\right) t_{\mf{g}/\mf{h}} + 2\pi i \left(\kappa + g^{\vee}_\mf{h}\right)t_{\mf{h}}}}{e^{-2\pi i(\Phi+w\tau,\rho_\mf{g})-\frac{2\pi i \tau}{2g_\mf{g}^{\vee}}\rho_\mf{g}^2} \sum_{\sigma \in W_{\mf{g}}}\epsilon(\sigma) \Theta_{\sigma(\rho_\mf{g})}^{g_\mf{g}^{\vee},\mf{g}}\left(\Phi+w\tau;\tau,t_{\mf{g}/\mf{h}},t_{\mf{h}}\right)},
\eqne
where
\eqnb
\Theta_{\mu}^{k,\mf{g}}\left(\Phi;\tau,t_{\mf{g}/\mf{h}},t_{\mf{h}}\right)
	&\equiv&
		e^{2\pi i k t_{\mf{g}/\mf{h}} + 2\pi i \kappa t_{\mf{h}}}\sum_{\beta\in L_\mf{g}^{\vee}} e^{2\pi i(\Phi,\mu + k\beta) + \frac{\pi i \tau}{k}\left(\mu + k\beta\right)^2}\,.
\eqne
Using the definition of the theta function and that $w$ lies in the coroot lattice one obtains
\eqnb
\Theta_{\mu}^{k,\mf{g}}\left(\Phi + w\tau;\tau,t_{\mf{g}/\mf{h}},t_{\mf{h}}\right)
	&=&
		e^{-2\pi i k(\Phi,w) - \pi i k \tau (w,w)}\Theta_{\mu}^{k,\mf{g}}\left(\Phi;\tau,t_{\mf{g}/\mf{h}},t_{\mf{h}}\right)\,,
\eqne
which inserted into the character yields
\eqnb
\chi_{\mu,w}^{k,\mf{g}}(\Phi;\tau,t_{\mf{g}/\mf{h}},t_{\mf{h}})
	&=&
		\frac{e^{\pi i \tau \frac{(\mu + \rho_\mf{g} + (k+g^{\vee}_\mf{g})w)^2}{(k+g_\mf{g}^{\vee})} + 2\pi i (\Phi,\mu + \rho_\mf{g} + (k+g_\mf{g}^{\vee})w) +  2\pi i \left(k + g_\mf{g}^{\vee}\right) t_{\mf{g}/\mf{h}} + 2\pi i \left(\kappa + g^{\vee}_\mf{h}\right)t_{\mf{h}}}}{\sum_{\sigma \in W_\mf{g}}\epsilon(\sigma) \Theta_{\sigma(\rho_\mf{g})}^{g_\mf{g}^{\vee},\mf{g}}\left(\Phi;\tau,t_{\mf{g}/\mf{h}},t_{\mf{h}}\right)}\,.
	\no
	\label{charGw}
\eqne

\section{Properties under modular transformations}
\label{sec:Properties}

In this section we will further extend the set of $\hat{\mf{g}}$ modules such that certain combinations will transform nicely under $SL(2,\Z)$, and can therefore serve as (chiral) building blocks for modular invariant bulk spectra.

\subsection{The $\mf{g}$ sector}

The characters (\ref{charGw}) still do not automatically transform nicely under modular transformations. However, the form of the numerator reveals that summing over $w\in L^\vee_\mf{g}$ as well as over all $\sigma(\mu+\rho_\mf{g})$, $\sigma\in W_\mf{g}$ with sign, gives a familiar combination of theta functions also in the numerator.
The result is
\eqnb
\tilde{\chi}_{\mu}^{k,\mf{g}}(\Phi;\tau,t_{\mf{g}/\mf{h}},t_{\mf{h}})
	&=&
		\frac{\sum_{\sigma \in W_\mf{g}}\epsilon(\sigma) \Theta_{\sigma(\mu + \rho_\mf{g})}^{k+g_\mf{g}^{\vee},\mf{g}}\left(\Phi;\tau,t_{\mf{g}/\mf{h}},t_{\mf{h}}\right)}{\sum_{\sigma \in W}\epsilon(\sigma) \Theta_{\sigma(\rho_\mf{g})}^{g^{\vee}_\mf{g},\mf{g}}\left(\Phi;\tau,t_{\mf{g}/\mf{h}},t_{\mf{h}}\right)}.
\label{Character:gw}
\eqne
Three remarks are in order concerning this expression.

First, it is clear that we have extended the set of $\mf{g}$-modules from $P_-^{k,\mf{g}}$ to 
\[ \cup_{\sigma\in W_\mf{g}}\sigma(P_-^{k,\mf{g}}+\rho_\mf{g}) - \rho_\mf{g}.\]
This set now no longer consists of only highest weight modules, but there are also lowest weight modules and modules with extremal but neither highest nor lowest weights (w.r.t.\ the chosen triangular decomposition of $\mf{g}$).

Second, we cannot justify the expression (\ref{Character:gw}) as a function of $\tau$ on the Teichm{\"u}ller space of the torus. Since $k+g_\mf{g}^\vee<0$ the summation over $w\in L_\mf{g}^\vee$ only converges when $\tau$ lies in the {\em lower} half plane, but in expressing the denominator as a combination of theta functions we have already chosen $\tau$ to lie in the upper half plane. This is of course not surprising, seeing that we are considering a sigma model with a target space of Lorentzian signature. Furthermore, restricting to integral weights we expect that the target space is not the universal cover of the $G/Ad(H)$ but rather contains closed timelike curves. Instead of a wrong signature Gaussian integral we are thus left with a wrong signature theta function. Note that the same happens for the $SL(2,\R)$ WZNW model on any finite cover of the group \cite{Henningson:1991jc}. The extended character (\ref{Character:gw}) should then be interpreted as a formal expression representing a regularised version. There is of course a risk that any regularisation of the numerator will spoil the modular properties we will soon determine. We do not address this issue here, but hope to return to it in the future. When handling expressions like (\ref{Character:gw}) one can treat the numerator and denominator as formal series in $q^{-1}$ respectively $q$ (where $q \equiv e^{2\pi i \tau}$), allowing only at the end of manipulations to cancel common factors.

Third, note that the extended character takes a form very reminiscent of the characters of irreducible integral dominant highest weight modules of $\hat{\mf{g}}$ at positive integer level, i.e.\ precisely those appearing in the compact WZNW model. Comparing the number of weights $\mu$ at level $k$ with the compact case one discovers that our expression corresponds to the compact model at level $-k-2g_\mf{g}^\vee$. This correspondence carries an intriguing flavour of the Kazhdan--Lusztig duality of the categories $\mathcal{O}_{|k|+g_\mf{g}^\vee}$ and $\mathcal{O}_{-(|k|+g_\mf{g}^\vee)}$ for the affine Lie algebra $\hat{\mf{g}}$ \cite{KazLusz}. In the present case, however, we are of course not dealing with integrable modules.

Let us now study properties of the extended character (\ref{Character:gw}) under modular transformations. Consider first the simplest one, the $\cT$-transformation. The theta functions transform as
\eqnb
\Theta_{\mu}^{k,\mf{g}}\left(\Phi;\tau+1,t_{\mf{g}/\mf{h}},t_{\mf{h}}\right)
	&=&
		e^{\frac{\pi i}{k}(\mu,\mu)}\Theta_{\mu}^{k,\mf{g}}\left(\Phi;\tau,t_{\mf{g}/\mf{h}},t_{\mf{h}}\right)
\eqne
from which one obtains
\eqnb
\tilde{\chi}_{\mu}^{k,\mf{g}}(\Phi;\tau+1,t_{\mf{g}/\mf{h}},t_{\mf{h}})
	&=&
		e^{2\pi i \left(\frac{(\mu+\rho_{\mf{g}})^2}{2(k+g_{\mf{g}}^{\vee})}-\frac{\rho^2_{\mf{g}}}{2g_{\mf{g}}^{\vee}}\right)}\,\tilde{\chi}_{\mu}^{k,\mf{g}}(\Phi;\tau,t_{\mf{g}/\mf{h}},t_{\mf{h}}).
	\no
	&=&
		\sum_{\nu \in P^{k,\mf{g}}_{-}}\cT^{k,\mf{g}}_{\mu,\nu}\,\tilde{\chi}_{\nu}^{k,\mf{g}}(\Phi;\tau,t_{\mf{g}/\mf{h}},t_{\mf{h}}),
\eqne
where
\eqnb
\cT^{k,\mf{g}}_{\mu,\nu}
	&=&
		e^{2\pi i \left(\frac{(\mu+\rho_{\mf{g}})^2}{2(k+g_{\mf{g}}^{\vee})}-\frac{\rho^2_{\mf{g}}}{2g_{\mf{g}}^{\vee}}\right)}\delta_{\mu,\nu}.
\eqne

Consider now the $\cS$-transformation. Using Poisson resummation one shows that the theta functions transform as 
\eqnb
&&\Theta_{\mu}^{k,\mf{g}}\left(\frac{\Phi}{\tau};-\frac{1}{\tau},t_{\mf{g}/\mf{h}}-\frac{\phi^2G^{\mf{g}}_{11}}{2\tau},t_{\mf{h}}-\frac{\left(\theta,\theta\right)_{\mf{h}}}{2\tau}\right)
\no
&&=
	\left(\frac{-i\tau}{k}\right)^{r_{\mf{g}}/2}\frac{1}{Vol(L_{\mf{g}}^{\vee})}\sum_{\nu\in P^{\mf{g}}/kL_{\mf{g}}^{\vee}}e^{-2\pi i \frac{(\nu,\mu)}{k}}\Theta_{\nu}^{k,\mf{g}}\left(\Phi;\tau,t_{\mf{g}/\mf{h}},t_{\mf{h}}\right).
\eqne
The summation over $P^\mf{g}/kL_\mf{g}^\vee$ can equivalently be described as a summation over weights in the weight lattice enclosed by the (fundamental) Brioullin zone of the sublattice $kL_\mf{g}^{\vee}$. To construct this zone we consider vectors pointing from a given point in the lattice, here chosen to be the origin, to the closest points in the lattice in all directions. The number of such vectors is the number of short coroots, and they can be represented as $\{k\alpha^{\vee}: \alpha\in \Delta_\mf{g}, \; (\alpha^{\vee},\alpha) = 2 \}$. The zone is then defined by the hyper volume enclosed by the surfaces orthogonal to the vectors located at half the distance from the origin. This zone depends on $k$ and we denote it by $\mathfrak{B}^\mf{g}_k$. Note that the distance between the origin and any one of the bounding surfaces is $\frac{k}{2}\left(\alpha,\alpha^{\vee}\right) = k$. We now have
\eqnb
\sum_{\nu\in P^\mf{g}/kL_\mf{g}^{\vee}} f(\nu)
	&=&
		\sum_{\nu \in \mathfrak{B}^\mf{g}_k }f(\nu).
\eqne
As all weights $\nu$ are obtained by Weyl reflections from weights in the fundamental Weyl chamber, we can rewrite the expression above as
\eqnb
\sum_{\nu \in \mathfrak{B}^\mf{g}_k }f(\nu)
	&=&
		\sum_{\nu^i\geq 0,\, \nu \in \mathfrak{B}^\mf{g}_k }\sum_{\sigma \in W_\mf{g}}f(\sigma(\nu)).
\eqne
In the fundamental Weyl chamber, there is only one surface limiting the length of the weights. As the only short coroot in the Weyl chamber is the highest root $\theta_{\mf{g}}^{\vee}$, the weights satisfy $(\nu,\theta_{\mf{g}})\leq k$ and it follows that
\eqnb
\sum_{\nu \in \mathfrak{B}^\mf{g}_k }f(\nu)
	&=&
		\sum_{\nu^i \geq 0,\, (\nu,\theta_{\mf{g}}) \leq k}\sum_{\sigma \in W_\mf{g}}f(\sigma(\nu)).
\eqne

We will first consider the denominator of (\ref{Character:gw}). Using the description above of $P^\mf{g}/g_\mf{g}^{\vee}L_\mf{g}^{\vee}$ one obtains
\eqnb
&&
\sum_{\sigma\in W_\mf{g}}\epsilon(\sigma)\Theta_{\sigma(\rho_\mf{g})}^{g_\mf{g}^{\vee},\mf{g}}\left(\frac{\Phi}{\tau};-\frac{1}{\tau},t_{\mf{g}/\mf{h}}-\frac{\phi^2G^{\mf{g}}_{11}}{2\tau},t_{\mf{h}}-\frac{\left(\theta,\theta\right)_{\mf{h}}}{2\tau}\right)
\no
	&&=
\left(\frac{-i\tau}{g_\mf{g}^{\vee}}\right)^{r_\mf{g}/2}\frac{1}{Vol(L_\mf{g}^{\vee})}\sum_{\sigma\in W_\mf{g}}\sum_{\sigma'\in W_\mf{g}}\epsilon(\sigma)\no
	&& \times\sum_{\nu^{i}\geq 0, (\nu,\theta_{\mf{g}})\leq g_\mf{g}^{\vee}}
e^{-2\pi i \frac{(\sigma'(\nu),\sigma(\rho_\mf{g}))}{g_\mf{g}^{\vee}}}\Theta_{\sigma'(\nu)}^{g_\mf{g}^{\vee},\mf{g}}\left(\Phi;\tau,t_{\mf{g}/\mf{h}},t_{\mf{h}}\right)
\no
	&&=
\left(\frac{-i\tau}{g_\mf{g}^{\vee}}\right)^{r_\mf{g}/2}\frac{1}{Vol(L_\mf{g}^{\vee})}\sum_{\nu^{i}\geq 1, (\nu,\theta_{\mf{g}})\leq g^{\vee}-1}
\sum_{\sigma\in W_\mf{g}}\epsilon(\sigma)e^{-2\pi i \frac{(\nu,\sigma(\rho_\mf{g}))}{g_\mf{g}^{\vee}}}\no
	&& \times\sum_{\sigma'\in W_\mf{g}}\epsilon(\sigma')\Theta_{\sigma'(\nu)}^{g_\mf{g}^{\vee},\mf{g}}\left(\Phi;\tau,t_{\mf{g}/\mf{h}},t_{\mf{h}}\right)\, .
\eqne
In the last line we have used that a weight with at least one component equal to zero will not contribute because of the summation over the Weyl group. This is also the case for weights on the affine boundary of the Weyl chamber. The only weight satisfying the conditions in the expression above is the Weyl vector, resulting in
\eqnb
&&
\sum_{\sigma\in W_\mf{g}}\epsilon(\sigma)\Theta_{\sigma(\rho_\mf{g})}^{g_\mf{g}^{\vee},\mf{g}}\left(\frac{\Phi}{\tau};-\frac{1}{\tau},t_{\mf{g}/\mf{h}}-\frac{\phi^2G^{\mf{g}}_{11}}{2\tau},t_{\mf{h}}-\frac{\left(\theta,\theta\right)_{\mf{h}}}{2\tau}\right)
\no
	&&=
\left(\frac{-i\tau}{g_\mf{g}^{\vee}}\right)^{r_\mf{g}/2}\frac{1}{Vol(L_\mf{g}^{\vee})}
\no
	&&\times
\sum_{\sigma\in W_\mf{g}}\epsilon(\sigma)e^{-2\pi i \frac{(\rho_\mf{g},\sigma(\rho_\mf{g}))}{g_\mf{g}^{\vee}}}\sum_{\sigma'\in W}\epsilon(\sigma')\Theta_{\sigma'(\rho_\mf{g})}^{g_\mf{g}^{\vee},\mf{g}}\left(\Phi;\tau,t_{\mf{g}/\mf{h}},t_{\mf{h}}\right)
\no
	&&=
(-i)^{d_\mf{g}/2}\tau^{r_\mf{g}/2}\sum_{\sigma\in W_\mf{g}}\epsilon(\sigma)\Theta_{\sigma(\rho_\mf{g})}^{g_\mf{g}^{\vee},\mf{g}}\left(\Phi;\tau,t_{\mf{g}/\mf{h}},t_{\mf{h}}\right).
\eqne
Next, consider the numerator of (\ref{Character:gw}). It is convenient to shift the weights to be summed over from $\nu$ to $\nu + \rho_{\mf{g}}$, modifying the restrictions on the sum to $\nu^i\geq 0$ and $\left(\nu,\theta_{\mf{g}}\right) \leq - k- 2g_{\mf{g}}^{\vee}$. Instead of summing over the fundamental Weyl chamber we would like to consider the reflections starting from the anti-fundamental Weyl chamber, where weights satisfy $(\nu,\alpha^{(i)})\leq 0$ for all simple roots. We do this by shifting the Weyl group element $\sigma$ to $\sigma\circ\sigma_{max}$, where $\sigma_{max}$ is the longest element in the Weyl group. This element is always a reflection, and thus $\epsilon(\sigma_{max}) = -1$. Evaluating the longest element on a general $\nu+\rho_{\mf{g}}$ one obtains $-\nu^{+}-\rho_{\mf{g}}$, where $\nu^{+}$ is the weight dual to $\nu$. Defining $\nu' = -\nu^{+} - 2\rho_{\mf{g}}$, the limits of the summation of the weights change from $\nu \geq 0$ to $\nu' \leq -2$ and the restriction $\left(\nu,\theta_{\mf{g}}\right) \leq -k-2g_{\mf{g}}^{\vee}$ to $\left(\nu',\theta_{\mf{g}}\right) \geq k + 2$ in the new variables. In this way we get
\eqnb
&&
\sum_{\sigma\in W_{\mf{g}}}\epsilon(\sigma)\Theta_{\sigma(\mu +\rho_{\mf{g}})}^{k+g_{\mf{g}}^{\vee}}\left(\frac{\Phi}{\tau};-\frac{1}{\tau},t_{\mf{g}/\mf{h}}-\frac{\phi^2G^{\mf{g}}_{11}}{2\tau},t_{\mf{h}}-\frac{\left(\theta,\theta\right)_{\mf{h}}}{2\tau}\right)
\no
	&&=
-\left(\frac{i\tau}{-(k+g_{\mf{g}}^{\vee})}\right)^{r_{\mf{g}}/2}\frac{1}{Vol(L_{\mf{g}}^{\vee})}
\sum_{\nu \in P_{-}^{k,\mf{g}}}\sum_{\sigma\in W}\epsilon(\sigma)e^{-2\pi i \frac{(\nu+\rho_{\mf{g}},\sigma(\mu+\rho_{\mf{g}}))}{k+g_{\mf{g}}^{\vee}}}
\no
	&&\times
\sum_{\sigma'\in W_{\mf{g}}}\epsilon(\sigma')\Theta_{\sigma'(\nu+\rho_{\mf{g}})}^{k+g_{\mf{g}}^{\vee}}\left(\Phi;\tau,t_{\mf{g}/\mf{h}},t_\mf{h}\right).
\eqne

The result of combining the numerator and denominator expressions for the extended character is
\eqnb
&&
\tilde{\chi}_{\mu}^{k}\left(\frac{\Phi}{\tau};-\frac{1}{\tau},t_{\mf{g}/\mf{h}}-\frac{\phi^2G^{\mf{g}}_{11}}{2\tau},t_{\mf{h}}-\frac{\left(\theta,\theta\right)_{\mf{h}}}{2\tau}\right)
\no
	&&=
-\frac{i^{\frac{d_\mf{g}-r_\mf{g}}{2}}}{i^{r_\mf{g}}\left[-(k+g_\mf{g}^{\vee})\right]^{r/2}} \frac{1}{Vol(L_\mf{g}^{\vee})}\sum_{\nu \in P_{-}^{k,\mf{g}}}\sum_{\sigma\in W_\mf{g}}\epsilon(\sigma)e^{-2\pi i \frac{\nu+\rho_\mf{g},\sigma(\mu+\rho_\mf{g}))}{k+g_\mf{g}^{\vee}}}\tilde{\chi}_{\mu}^{k}(\Phi;\tau,t_{\mf{g}/\mf{h}},t_\mf{h})
\no
	&&=
\sum_{\nu \in P_{-}^{k,\mf{g}}} \cS^{k,\mf{g}}_{\mu,\nu}\, \tilde{\chi}_{\nu}^{k}\left(\Phi;\tau,t_{\mf{g}/\mf{h}},t_\mf{h}\right),\label{Stransf:gw}
\eqne
where we have defined the matrix $\cS^{k,\mf{g}}_{\mu,\nu}$
\eqnb
\cS^{k,\mf{g}}_{\mu,\nu}
	&=&
		-\frac{i^{\frac{d_\mf{g}-r_\mf{g}}{2}}}{i^{r_\mf{g}}\left[-(k+g_\mf{g}^{\vee})\right]^{r_\mf{g}/2}} \frac{1}{Vol(L_\mf{g}^{\vee})}\sum_{\sigma\in W_\mf{g}}\epsilon(\sigma)e^{-2\pi i \frac{(\nu+\rho_\mf{g},\sigma(\mu+\rho_\mf{g}))}{k+g_\mf{g}^{\vee}}}.
\eqne
Note that the only point where we are, unjustifiably, identifying the modular parameter $\tau$ in the numerator with that in the denominator is when cancelling the factors of $\tau^{r_\mf{g}/2}$ resulting in (\ref{Stransf:gw}). Keeping the modular parameters of the numerator and denominator separate, the result is modified by an overall multiplicative factor equal to the ratio of the modular parameters to the power $r_\mf{g}/2$.

It is interesting to note that this matrix satisfies the following relation
\eqnb
\cS_{-\mu-2\rho_\mf{g},-\nu-2\rho_\mf{g}}^{-k-2g_\mf{g}^{\vee},\mf{g}}
	&=&
		-i^{d_\mf{g}}\left(\cS_{\mu,\nu}^{k,\mf{g}}\right)^{*}\,,\label{SmatrRel}
\eqne
where on the left hand side is written the Kac--Peterson modular $\cS$-matrix for a compact WZNW model at positive integer level $-(k+2g_\mf{g}^\vee)$ \cite{Kac:1984mq}. Again we see something akin to the Kazhdan--Lusztig duality of categories $\mathcal{O}$. The relation (\ref{SmatrRel}) immediately implies that our $\cS$ matrix is unitary.

\subsection{The other sectors}
Consider first the effect of spectral flow for the $\tilde{\mf{h}}$ and coset ghost sector. Using
\eqnb
\Theta^{\tilde{\kappa},\mf{h}}_{\tilde{\mu}}\left(\theta-w\tau;\tau,t_\mf{h}\right)
&=&
e^{-2\pi i \tilde{\kappa}(\theta,w) - \pi i \tilde{\kappa} \tau (w,w)}\Theta^{\tilde{\kappa},\mf{h}}_{\tilde{\mu}}\left(\theta;\tau,t_\mf{h}\right)
\eqne
one easily deduces that the modules considered are invariant under spectral flow, although the states within the modules are permuted.

The characters of the other sectors, (\ref{charH} - \ref{charVIR}), are
\eqnb
\chi^{\tilde{\kappa}}_{\tilde{\mu}}\left(\theta;\tau,t_\mf{h}\right)
	&=&
		\frac{\sum_{\sigma\in W_\mf{h}}\epsilon(\sigma)\Theta_{\sigma\left(\tilde{\mu}+\rho_\mf{h}\right)}^{\tilde{\kappa}+g_\mf{h}^{\vee}}\left(\theta;\tau,t_\mf{h}\right)}{\sum_{\sigma\in W_\mf{h}}\epsilon(\sigma)\Theta_{\sigma\left(\rho_\mf{h}\right)}^{g_\mf{h}^{\vee}}\left(\theta;\tau,t_\mf{h}\right)}
\\
\chi^{\mf{g}/\mf{h}\;gh}\left(\theta;\tau,t_\mf{h}\right)
	&=&
		\left(\sum_{\sigma\in W_\mf{h}}\epsilon(\sigma)\Theta_{\sigma(\rho_\mf{h})}^{g_\mf{h}^{\vee}}(\theta;\tau,t_\mf{h})\right)^2
\\
\chi^{\mathrm{Vir}\;gh}(\tau)
	&=&
		\eta^2(\tau),
\eqne
while the additional unitary CFT is assumed to possess characters with known modular properties. Under a $\cT$ transformation, these characters transform as
\eqnb
\chi^{\tilde\kappa,\tilde{\mf{h}}}_{\tilde{\mu}}\left(\theta;\tau+1,t_{\mf{h}}\right)
	&=&
		\sum_{\tilde{\nu} \in P^{\tilde{\kappa},\mf{h}}_{+}}\cT^{\tilde{\mf{h}}}_{\tilde{\mu},\tilde{\nu}}\chi^{\tilde{\kappa},\tilde{\mf{h}}}_{\tilde{\nu}}\left(\theta;\tau,t_{\mf{g}/\mf{h}},t_{\mf{h}}\right)
	\no
\chi^{\mf{g}/\mf{h}\; gh}\left(\theta;\tau+1,t_{\mf{h}}\right)
	&=&
		\cT^{\mf{g}/\mf{h}\; gh}\chi^{\mf{g}/\mf{h}\; gh}\left(\theta;\tau,t_{\mf{h}}\right)
	\no
\chi^{\mathrm{Vir}\; gh}\left(\tau+1\right)
	&=&
		\cT^{\mathrm{Vir}\; gh}\chi^{\mathrm{Vir}\; gh}\left(\tau\right)
	\no
\chi^{\mathrm{CFT}}_{\lambda}\left(\tau+1\right)
	&=&
		\sum_{\lambda'}\cT^{\mathrm{CFT}}_{\lambda,\lambda'}\chi^{\mathrm{CFT}}_{\lambda'}\left(\tau\right)
\eqne
where
\eqnb
\cT^{\tilde{\mf{h}}}_{\tilde{\mu},\tilde{\nu}}
	&=&
		e^{2\pi i \left(\frac{\left(\tilde{\mu} + \rho_{\mf{h}},\tilde{\mu} + \rho_{\mf{h}}\right)}{2\left(\tilde{\kappa}+g_{\mf{h}}^{\vee}\right)} - \frac{\left(\rho_{\mf{h}},\rho_{\mf{h}}\right)}{2g_{\mf{h}}^{\vee}}\right)}\delta_{\tilde{\mu},\tilde{\nu}}
	\no
\cT^{\mf{g}/\mf{h}\; gh}
	&=&
		e^{2\pi i \frac{\left(\rho_{\mf{h}},\rho_{\mf{h}}\right)}{g_{\mf{h}}^{\vee}}}
	\no
\cT^{\mathrm{Vir}\; gh}
	&=&
		e^{2\pi i \frac{1}{12}}
	\no
\cT^{\mathrm{CFT}}_{\lambda,\lambda'}
	&=&
		e^{2\pi i m_{\lambda}}\delta_{\lambda,\lambda'} \;
\eqne
The $\cS$ transformation has the following effect
\eqnb
\chi^{\tilde\kappa,\tilde{\mf{h}}}_{\tilde{\mu}}\left(-\frac{\theta}{\tau};-\frac{1}{\tau},t_{\mf{h}}-\frac{\left(\theta,\theta\right)_{\mf{h}}}{2\tau}\right)
	&=&
		\sum_{\tilde{\nu} \in P^{\tilde{\kappa},\mf{h}}_{+}}\cS^{\tilde{\mf{h}}}_{\tilde{\mu},\tilde{\nu}}\chi^{\tilde{\kappa},\tilde{\mf{h}}}_{\tilde{\nu}}\left(\theta;\tau,t_{\mf{h}}\right)
	\no
\chi^{\mf{g}/\mf{h}\; gh}\left(-\frac{\theta}{\tau};-\frac{1}{\tau},t_{\mf{h}}-\frac{\left(\theta,\theta\right)_{\mf{h}}}{2\tau}\right)
	&=&
		\tau^{r_{\mf{h}}}\cS^{\mf{g}/\mf{h}\; gh}\chi^{\mf{g}/\mf{h}\; gh}\left(\theta;\tau,t_{\mf{h}}\right)
	\no
\chi^{\mathrm{Vir}\; gh}\left(-\frac{1}{\tau}\right)
	&=&
		\tau \cS^{\mathrm{Vir}\; gh}\chi^{\mathrm{Vir}\; gh}\left(\tau\right)
	\no
\chi^{\mathrm{CFT}}_{\lambda}\left(-\frac{1}{\tau}\right)
	&=&
		\sum_{\lambda'}\cS^{\mathrm{CFT}}_{\lambda,\lambda'}\chi^{\mathrm{CFT}}_{\lambda'}\left(\tau\right)
\eqne
where
\eqnb
\cS^{\tilde{\mf{h}}}_{\tilde{\mu},\tilde{\nu}}
	&=&
		\frac{ i^{ \frac{ d_{ \mf{h} } - r_{ \mf{h} } }{2} } }{\left(\tilde{\kappa} - g^{\vee}_{\mf{h}}\right)^{r/2}} \frac{1}{Vol(L_{\mf{h}}^{\vee})} \sum_{\sigma \in W_\mf{h}}\epsilon(\sigma)e^{-2\pi i \frac{\left(\tilde{\mu} + \rho_{\mf{h}},\tilde{\mu} + \rho_{\mf{h}}\right)}{\tilde{\kappa} + g_{\mf{h}}^{\vee}}}
	\no
\cS^{\mf{g}/\mf{h}}
	&=&
		(-i)^{d_{\mf{h}}}
	\no
\cS^{\mathrm{Vir}\; gh}
	&=&
		-i,
\eqne
and $S^{\mathrm{CFT}}_{\lambda,\lambda'}$ is a unitary matrix.

For the combined character (\ref{fullchardecom}) with $\chi_{\mu}^{k,\mf{g}}$ replaced by the extended character (\ref{Character:gw}) we thus have
\eqnb
\chi^{k}_{\mu,\tilde{\mu},\lambda}(\Phi;\tau+1,t_{\mf{g}/\mf{h}})
	&=&
		\sum_{\nu,\tilde{\nu},\lambda'}\cT_{\{\mu,\tilde{\mu},\lambda\},\{\nu,\tilde{\nu},\lambda'\}}\chi^{k}_{\nu,\tilde{\nu},\lambda}(\Phi;\tau,t_{\mf{g}/\mf{h}})
	\\
\chi^{k}_{\mu,\tilde{\mu},\lambda}\left(\frac{\Phi}{\tau};-\frac{1}{\tau},t_{\mf{g}/\mf{h}}-\frac{\phi^2G^{\mf{g}}_{11}}{2\tau}\right)
	&=&
		\tau^{r_\mf{g}}\sum_{\nu,\tilde{\nu},\lambda'}\cS_{\{\mu,\tilde{\mu},\lambda\},\{\nu,\tilde{\nu},\lambda'\}}\chi^{k}_{\nu,\tilde{\nu},\lambda}(\Phi;\tau,t_{\mf{g}/\mf{h}})\, ,
\eqne
where 
\eqnb
\cT_{\{\mu,\tilde{\mu},\lambda\},\{\nu,\tilde{\nu},\lambda'\}}
	&=&
		\cT^{\mf{g}}_{\mu,\nu}\cT^{\tilde{\mf{h}}}_{\tilde{\mu},\tilde{\nu}}\cT^{\mf{g}/\mf{h}\; gh}\cT^{\mathrm{Vir}\; gh} \cT^{CFT}_{\lambda,\lambda'}
	\\
\cS_{\{\mu,\tilde{\mu},\lambda\},\{\nu,\tilde{\nu},\lambda'\}}
	&=&
		\cS^{\mf{g}}_{\mu,\nu}\cS^{\tilde{\mf{h}}}_{\tilde{\mu},\tilde{\nu}}\cS^{\mf{g}/\mf{h}\; gh}\cS^{\mathrm{Vir}\; gh} \cS^{CFT}_{\lambda,\lambda'}.
\eqne
In the formulas above, as well as henceforth, summations over $\mu$, $\nu$ respectively $\tilde{\mu}$, $\tilde{\nu}$ are implicitly assumed to have ranges $P_-^{k,\mf{g}}$ respectively $P_+^{\tilde{\kappa},\mf{h}}$.

\section{Modular invariant partition functions}
\label{sec:ModularPart}

Define the combination
\eqnb
B^{k,\mf{g},\mf{h}}\left(\Phi;\tau,t_{\mf{g}/\mf{h}}\right)
	&\equiv&
		\sum_{\mu,\tilde{\mu},\lambda} \left(\chi^{k}_{\mu,\tilde{\mu},\lambda}(\Phi;\tau,t_{\mf{g}/\mf{h}})\right)^{*}\chi^{k}_{\mu,\tilde{\mu},\lambda}(\Phi;\tau,t_{\mf{g}/\mf{h}}).
\eqne
Since all matrices $\cT$ are manifestly unitary, this function is trivially invariant under a $\cT$ transformation. Consider therefore an $\cS$ transformation of this function
\eqnb
B^{k,\mf{g},\mf{h}}\left(\frac{\Phi}{\tau};-\frac{1}{\tau},t_{\mf{g}/\mf{h}}-\frac{\phi^2G^{\mf{g}}_{11}}{2\tau}\right)
	&=&
		\left|\tau\right|^{2r_{\mf{g}}}\sum_{\mu,\tilde{\mu},\lambda} \sum_{\mu',\tilde{\mu}',\lambda'} \sum_{\mu'',\tilde{\mu}'',\lambda''}
	\no
	&\times&
		\left(\cS_{\{\mu,\tilde{\mu},\lambda\},\{\mu',\tilde{\mu}',\lambda'\}}\right)^{\dagger}\cS_{\{\mu,\tilde{\mu},\lambda\},\{\mu'',\tilde{\mu}'',\lambda''\}}
	\no
	&\times&
		\left(\chi^{k}_{\mu',\tilde{\mu}',\lambda'}(\Phi;\tau,t_{\mf{g}/\mf{h}})\right)^{*}\chi^{k}_{\mu'',\tilde{\mu}'',\lambda''}(\Phi;\tau,t_{\mf{g}/\mf{h}})
	\no
	&=&
		\left|\tau\right|^{2r_{\mf{g}}}\sum_{\mu,\tilde{\mu},\lambda}
		\left(\chi^{k}_{\mu,\tilde{\mu},\lambda}(\Phi;\tau,t_{\mf{g}/\mf{h}})\right)^{*}\chi^{k}_{\mu,\tilde{\mu},\lambda}(\Phi;\tau,t_{\mf{g}/\mf{h}})
	\no
	&=&
		\left|\tau\right|^{2r_{\mf{g}}} B^{k,\mf{g},\mf{h}}\left(\Phi;\tau,t_{\mf{g}/\mf{h}}\right).
\eqne
Note that $\left|\tau\right|^{2r_{\mf{g}}}$ can be divided into two contributions, $\left|\tau\right|^{2r_{\mf{h}}}$ and $\left|\tau\right|^{2}$, where the first arises from the coset ghosts and the second from the Virasoro ghosts. Recall that we must integrate over $\theta$ and $\tau$ to project down to physical string states. Let us first consider the parameter $\theta$.
The integration measure for $\theta$ and $\bar{\theta}$ transforms as
\eqnb
d^{r_{\mf{h}}}\theta d^{r_{\mf{h}}}\bar{\theta}
	&\stackrel{\cT}{\mapsto}&
		d^{r_{\mf{h}}}\theta d^{r_{\mf{h}}}\bar{\theta} 
	\no
d^{r_{\mf{h}}}\theta d^{r_{\mf{h}}}\bar{\theta}
	&\stackrel{\cS}{\mapsto}&
		\frac{d^{r_{\mf{h}}}\theta d^{r_{\mf{h}}}\bar{\theta}}{\left|\tau\right|^{2r_{\mf{h}}}}.
\eqne
Disregarding for a moment the Virasoro ghosts, the combination
\[d^{r_{\mf{h}}}\theta d^{r_{\mf{h}}}\bar{\theta}B^{k,\mf{g},\mf{h}}\left(\Phi;\tau,t_{\mf{g}/\mf{h}}\right)\]
is modular invariant, and the expression
\eqnb
\widetilde{B}(\phi;\tau,t_{\mf{g}/\mf{h}}) & \equiv & \int d^{r_{\mf{h}}}\theta d^{r_{\mf{h}}}\bar{\theta}B^{k,\mf{g},\mf{h}}\left(\Phi;\tau,t_{\mf{g}/\mf{h}}\right)\label{diagpnfn}
\eqne
is a candidate for a modular invariant partition function for a non-compact coset CFT. Since the Virasoro ghosts are included, however, we cannot expect to get a truly modular invariant partition function. Rather we should consider the combination
\eqnb
\frac{d\tau d\bar{\tau}}{\mathrm{im}(\tau)}d^{r_{\mf{h}}}\theta d^{r_{\mf{h}}}\bar{\theta}\,B^{k,\mf{g},\mf{h}}\left(\Phi;\tau,t_{\mf{g}/\mf{h}}\right).\label{diagmodinv}
\eqne
Since the measure of the moduli space for the torus satisfies
\eqnb
\frac{d\tau d\bar{\tau}}{\mathrm{im}(\tau)}
	&\stackrel{\cT}{\mapsto}&
		\frac{d\tau d\bar{\tau}}{\mathrm{im}(\tau)}
	\no
\frac{d\tau d\bar{\tau}}{\mathrm{im}(\tau)}
	&\stackrel{\cS}{\mapsto}&
		\frac{d\tau d\bar{\tau}}{\mathrm{im}(\tau)}\frac{1}{\left|\tau \right|^2},
\eqne
the expression (\ref{diagmodinv}) is truly modular invariant. Integrating (\ref{diagmodinv}) over the moduli space of a once punctured torus we get a well-defined torus string amplitude.

In general we can also find other modular invariant combinations. For any modular invariant $Z^\mf{g}_{\mu,\nu}$ of the compact $G$ WZNW model at level $-k-2g^\vee_{\mf{g}}$ and any modular invariants $Z^\mf{h}_{\tilde{\mu},\tilde{\nu}}$, $Z^{CFT}_{\lambda,\lambda'}$ of the $\widetilde{H}$ model at level $\tilde{\kappa}$ respectively the additional unitary CFT, we can define a well-defined torus string amplitude by coupling chiral and anti-chiral sectors using
\[Z_{\mu\tilde{\mu}\lambda,\nu\tilde{\nu}\lambda'}\equiv Z^\mf{g}_{\mu,\nu}Z^\mf{h}_{\tilde{\mu},\tilde{\nu}}Z^{CFT}_{\lambda,\lambda'}.\]
Due to the projection down to physical states, however, it is possible that inequivalent choices of modular invariants in the respective sectors still gives rise to equivalent string amplitudes.

\section{Discussion}
\label{sec:disc}

Let us briefly summarise what we have done so far. We used the construction \cite{Bjornsson:2007ha} of a class of non-compact $G/Ad(H)$ coset string models with unitary string spectrum. For any model in this class we started out by considering a subset of the set of $\hat{\mf{g}}$-modules allowed by unitarity, extended this set by the action of spectral flow together with the finite Weyl group of $\mf{g}$. Using the resulting modules of $\hat{\mf{g}}$ we considered blocks related by spectral flow and the finite Weyl group, and showed that characters of these blocks (the extended characters (\ref{Character:gw})) behave nicely under modular transformations. Combined characters of all sectors of the model then also transform in a well behaved manner, and we showed that any set of modular invariant torus partition functions of the individual sectors gives a modular invariant density on the moduli space of a once punctured torus that can in turn be integrated to yield a torus string amplitude.

As remarked in section \ref{sec:Properties} what we call extended characters of the $\mf{g}$ sector are really not well-defined functions of the modulus $\tau$. The reason is that summing over spectral flow yields a theta function that is convergent in the lower half plane, while the rest of the characters are defined in the upper half plane. Note that a similar problem appears already in the $SL(2,\R)$ WZNW model on a finite cover \cite{Henningson:1991jc} where it is still not resolved, as well as on the infinite cover \cite{Maldacena:2000hw} where it can be understood as an analytic continuation of the partition function of the so called $SL(2,\C)/SU(2)$ WZNW model. Clearly this is a crucial point that, at the moment,  prevents us from claiming that our construction gives well-defined string amplitudes. Nevertheless, we are optimistic and hope that it will be possible to find a regularisation preserving the modular properties.

Recall that in rational CFT we demand certain properties of the torus partition function other than just modular invariance. In particular it is required that the closed string vacuum sector appears exactly once in the bulk spectrum, implying $Z_{0,0}=1$. In none of our models does the trivial $\mf{g}$-module appear, and this is also not necessary for a modular invariant spectrum. Therefore even the modular invariant (\ref{diagpnfn}) does not possess a straightforward interpretation as a charge conjugation (or, depending on conventions, diagonal) modular invariant. One can speculate whether the $\hat{\mf{g}}$-module with highest weight $\mu$ closest to the tip of the anti-fundamental Weyl chamber, or rather the spectral flow \& Weyl group extended block corresponding to this module, plays a role similar to the vacuum block in a rational CFT. The motivation for this speculation comes from the relation between the non-compact theory at level $k$ and the compact theory at level $-k-2g^\vee_\mf{g}$ where this module corresponds to the identity in the compact case. Tensoring with this module, however, will in general yield a very complicated decomposition, unlikely to be truncated to the identity in the fusion product.

Apart from the question of regularising the extended characters there is from the point of view of string theory one additional important question. Since, if we interpret the coset model as a sigma model, the target space will have closed timelike curves, it appears necessary to unwrap some compact directions by going to a suitable covering space. From a representation theoretic perspective such a procedure is likely to make the unitary zero-mode spectrum denser.
A hint of which compact directions to unwrap is obtained from the proof of sufficient as well as necessary conditions for unitarity \cite{Bjornsson:2007ha,Bjornsson:2009dp}. Here all but the first component of the weights have to be integers, and therefore a natural guess is that only one direction needs to be unwrapped. On the $n$th cover of the compact direction, the first component of the weights can also adopt rational values, $\mu^1\in \Z/n$. As $(w^L,\mu)\in \Z$ and $(w^R,\mu)\in \Z$, this will enforce $w_1^L, w_1^R \in n\Z$, thus making the allowed spectral flow parameters less dense in the coroot lattice. On the $n$th cover there also appears additional possibilities with $w_1^L - w_1^R = 0\; \mathrm{Mod}(n)$, coupling chiral and anti-chiral sectors. Considering the $\mf{g}$ sector, the contribution from the $\mu$-block in the ``charge conjugation''  partition function would then change to
\eqnb
&&
\hspace{-1.3cm}
\sum_{w^1 = -[n/2]}^{n-[n/2]-1} \left[\sum_{w_1^R\in n\Z}\sum_{w^R_\mf{h}\in \Z}\sum_{\sigma^R\in W_\mf{g}}\epsilon(\sigma^R)\left(\chi^{k,\mf{g}}_{\sigma^R(\mu),w+w^R}\right)^{*}
\right.
\no
&&\times
\left.
\sum_{w_1^L\in n\Z}\sum_{w^L_\mf{h}\in \Z}\sum_{\sigma^L\in W_\mf{g}}\epsilon(\sigma^L)\chi^{k,\mf{g}}_{\sigma^L(\mu),w+w^L}\right]\,,
\label{finitecover}
\eqne
where $[\cdot]$ is the integer part and $w_\mf{h}$ is the $\mf{h}$ part of $w$. A way to get to the infinite cover is to take the naive limit $n\rightarrow \infty$ of (\ref{finitecover}), resulting in
\eqnb
\sum_{w^1 \in \Z} \left[\sum_{w^R_\mf{h}\in \Z}\sum_{\sigma^R\in W_\mf{g}}\epsilon(\sigma^R)\left(\chi^{k,\mf{g}}_{\sigma^R(\mu),w+w^R}\right)^{*}\sum_{w^L_\mf{h}\in \Z}\sum_{\sigma^L\in W_\mf{g}}\epsilon(\sigma^L)\chi^{k,\mf{g}}_{\sigma^L(\mu),w+w^L}\right]\, ,
\eqne
where $\mu^1$ in the limit can take a continuous set of values.

Let us briefly discuss the coset CFT one obtains when gauging the maximally compact subalgebra of $\mf{g}$ (without applying the Virasoro constraints). In contrast to the $\mf{h}$-gauging, there will be an additional $\hat{\mf{u}}_{\tilde{\kappa}}(1)$ in the auxiliary sector and the corresponding ghosts. The construction of such a model will then follow analogously to the construction in this paper, and one can form characters, modified by a factor $e^{-\pi i {c\tau}/{12}}$, which transform linearly under modular transformations. These can be used to obtain modular invariant partition functions, one example being
\eqnb
\int d\phi d\bar{\phi} d^{r_\mf{h}}\theta d^{r_\mf{h}}\bar{\theta}\,\sum_{\mu,\tilde{\mu},\tilde{\nu}}\,\left(\chi^{k}_{\mu,\tilde{\mu},\tilde{\nu}}\right)^{*}\chi^{k}_{\mu,\tilde{\mu},\tilde{\nu}},
\eqne
where $\tilde{\nu}$ is the weight of the auxiliary $\hat{\mf{u}}_{\tilde{\kappa}}(1)$.

It is interesting to compare the character of the irreducible module with highest weight $\mu$ of the compact affine Lie algebra at a given level $k$ with the extended character of the block corresponding to the modules of extremal weights $\sigma(-\mu-\rho)-\rho$, $\sigma\in W$ together with their images under spectral flow of the non-compact affine algebra at level $-k-2g^\vee$. In the compact case, the character of an irreducible module with dominant integral highest weight $\mu$ is constructed by adding and removing Verma modules with highest weights $\sigma(\mu+\rho) -\rho + (k+g^{\vee})\alpha^{\vee}$ where $\sigma\in W$ and $\alpha^\vee\in L^\vee$. These weights are in one to one correspondence with the modules we consider in the non-compact case generated by the extremal weights $\sigma(-\mu-\rho)-\rho$, $\sigma\in W$ together with their images under spectral flow. As the the two sets coincide and the modular properties of the characters respectively extended characters are related as shown in (\ref{SmatrRel}), it is an indication that the two are related, which in turn might be related to the Kazhdan--Lusztig duality.

We have not in this paper not considered the modules of $\hat{\mf{g}}$ constructed from unitary $\mf{g}$-modules \cite{Jakobsen:1983,Enright:1983}. With an extension similar to the one used in this paper we conjecture that such a starting point will give the same models as discussed here. Let us summarise the indications that this is indeed true. Consider the character for the $\hat{\mf{g}}$ sector first, which can be obtained by a straightforward generalisation of the character derived for $\mf{g}$ in \cite{jonas}. The character involves addition and subtraction by Verma modules with highest weights reached by Weyl reflections for the finite dimensional $\mf{h}$ subalgebra. Apart from these highest weight modules, one can also consider lowest and mixed extremal weights, which are connected by elements in the Weyl group in $\mf{g}$ but not $\mf{h}$. It is conceivable that from these extremal highest weight modules one can generate additional modules by using spectral flow, which is generated by elements in the coroot lattice. Defining extended characters by the blocks corresponding to these modules, it is likely that one obtains the same extended characters as presented here. The part in need of more work is $\tilde{\mf{h}}$. The main difficulty is that we do not know the characters of the modules one starts with. In this case, only the finite dimensional part is known since the highest weights are anti-dominant. As the affine part is dominant, it is difficult to determine which Verma modules are embedded. One could, however, argue that one needs to construct extended characters for this sector as well by adding lowest and mixed extremal weight modules generated by Weyl reflections of the anti-dominant highest weight modules, as well as spectral flowed sectors generated by a subset of the coroot lattice. It is then likely that the extended characters coincide with those of the integrable modules of $\hat{\mf{h}}$. Assuming that one can show that the physical spectrum is unitary, the combined characters would be identical to the ones obtained in this paper.

\vspace{1cm}
\noindent
{\bf{Acknowledgements}}
JB would like to thank the theory group at Karlstad University and Nanjing University for the hospitality during the completion of this work. JB acknowledges the support of the Swedish Research Council under project no.\ 623-2008-7048. JF is supported in parts by NSFC grant no.\ 10775067 as well as the Research Links Programme of Swedish Research Council under contract no.\ 348-2008-6049. JF would like to thank the theory group at Karlstad University where part of this work was carried out.


\begin{thebibliography}{99}

\bibitem{Bardakci:1970nb}
  K.~Bardakci and M.~B.~Halpern,
  Phys.\ Rev.\  D {\bf 3} (1971) 2493.\\
  M.~B.~Halpern,
  Phys.\ Rev.\  D {\bf 4} (1971) 2398.


\bibitem{Goddard:1984vk}
  P.~Goddard, A.~Kent and D.~I.~Olive,
  Phys.\ Lett.\  B {\bf 152} (1985) 88.

\bibitem{Bjornsson:2007ha}
  J.~Bj{\"o}rnsson and S.~Hwang,
   Nucl.\ Phys.\  B {\bf 797} (2008) 464, { [arXiv:0710.1050 [hep-th]]}.

\bibitem{Bjornsson:2008fq}
  J.~Bj\"ornsson and S.~Hwang,
  Nucl.\ Phys.\  B {\bf 812} (2009) 525
  [arXiv:0802.3578 [hep-th]].
  
\bibitem{Bjornsson:2009dp}
  J.~Bj\"ornsson and S.~Hwang,
  Nucl.\ Phys.\  B {\bf 832} (2010) 52
  [arXiv:0911.0264 [hep-th]].

\bibitem{Bjornsson:2010ja}
  J.~Bj\"ornsson and S.~Hwang,
  [arXiv:1001.3596 [hep-th]].

\bibitem{Karabali:1989dk}
  D.~Karabali and H.~J.~Schnitzer,
  Nucl.\ Phys.\  B {\bf 329} (1990) 649.

\bibitem{Hwang:1990aq}
  S.~Hwang,
  Nucl.\ Phys.\  B {\bf 354} (1991) 100.
  
\bibitem{Henningson:1991jc}
  M.~Henningson, S.~Hwang, P.~Roberts and B.~Sundborg,
  Phys.\ Lett.\  B {\bf 267} (1991) 350.
  
\bibitem{Maldacena:2000hw}
  J.~M.~Maldacena and H.~Ooguri,
  J.\ Math.\ Phys.\  {\bf 42} (2001) 2929
  [arXiv:hep-th/0001053].

\bibitem{Nappi:1993ie}
  C.~R.~Nappi and E.~Witten,
  Phys.\ Rev.\ Lett.\  {\bf 71} (1993) 3751
  [arXiv:hep-th/9310112].

\bibitem{Kiritsis:1993jk}
  E.~Kiritsis and C.~Kounnas,
  Phys.\ Lett.\  B {\bf 320} (1994) 264
  [Addendum-ibid.\  B {\bf 325} (1994) 536]
  [arXiv:hep-th/9310202].

\bibitem{Chen:2009wh}
  G.~Chen, Y.~K.~Cheung, Z.~Fan, J.~Fjelstad and S.~Hwang,
  Phys.\ Rev.\  D {\bf 80} (2009) 086003
  [arXiv:0907.5366 [hep-th]].

\bibitem{Gawedzki:1991yu}
  K.~Gawedzki,
  published in NATO ASI: Cargese 1991:0247-274,
  [arXiv:hep-th/9110076].
  
\bibitem{Fuchs:1997jv}
  J.~Fuchs and C.~Schweigert,
  ``Symmetries, Lie Algebras And Representations: A Graduate Course For
  Physicists,''
  {\it  Cambridge, UK: Univ. Pr. (1997) 438 p.}
  
\bibitem{Jakobsen:1983}
  H.~Jakobsen,
  J.\ Funct.\ Anal.\ {\bf 52} (1983) 385.
  
\bibitem{Hwang:1993nc}
  S.~Hwang and H.~Rhedin,
  Nucl.\ Phys.\  B {\bf 406} (1993) 165.
  {[arXiv:hep-th/9305174]}.

\bibitem{KazLusz}
D. Kazhdan and G. Lusztig,
J. Amer. Math. Soc. 6 (1993), 905-947\\
D. Kazhdan and G. Lusztig,
J. Amer. Math. Soc. 6 (1993), 949-1011\\
D. Kazhdan and G. Lusztig,
J. Amer. Math. Soc. 7 (1994), 335-381\\
D. Kazhdan and G. Lusztig,
J. Amer. Math. Soc. 7 (1994), 383-453.

\bibitem{Kac:1984mq}
  V.~G.~Kac and D.~H.~Peterson,
  Adv.\ Math.\  {\bf 53} (1984) 125.

\bibitem{Enright:1983}
  T.~Enright, R.~Howe and N.~Wallach,
  Progr.\ Math.\ 40:97--143 (1983), Birkh\"auser.

\bibitem{jonas}
  J.\ Bj{\"o}rnsson, ``Strings, Branes and Non-trivial Space-times,''
  Karlstad University Studies 2008:20

  	
\end{thebibliography}
\end{document}